\documentclass[aps,twocolumn,showpacs,amssymb]{revtex4}

\usepackage[english]{babel}
\usepackage{amssymb}
\usepackage{indentfirst}
\usepackage{graphics}
\usepackage{amsmath}
\usepackage{amsthm}

\usepackage{CatEn}

\usepackage{amsfonts,epsfig,color,bbm}

\begin{document}

\title{Algorithmic complexity and entanglement of quantum states}

\author{Caterina E. Mora$^1$ and Hans J. Briegel$^{1,2}$}

\affiliation{$^1$ Institut f\"ur Quantenoptik und Quanteninformation
  der \"Osterreichischen Akademie der Wissenschaften, Innsbruck,
  Austria\\
$^2$ Institut f{\"u}r Theoretische Physik, Universit{\"a}t Innsbruck,
Technikerstra{\ss}e 25, A-6020 Innsbruck, Austria} 

\date{\today}

\begin{abstract}
We define the algorithmic complexity of a quantum state
relative to a given precision parameter, and give upper bounds for
various examples of states. We also establish a connection between the
entanglement of a quantum state and its algorithmic complexity.
\end{abstract}
\maketitle

Algorithmic information provides a concise notion of
complexity, or randomness, for individual classical objects. It
is measured by the 
length of the shortest computer program that produces a faithful image
of the object, usually represented as a string of binary numbers
\cite{solomonoff,kolmogorov,chaitin}. This notion of complexity has
not only added a new perspective to abstract areas such as
mathematical proof theory (G\"odel's theorem) \cite{chaitin} but it
has also been 
applied successfully to a range of problems in
thermodynamics. Examples are the Maxwell demon paradox \cite{bennett,
  zurek}, and the treatment of irreversibility in classical chaotic systems \cite{schack, caves}.
 
Quantum theory has changed our
conception of physical objects, whose states are described as
vectors in a Hilbert space. For composite objects, this
leads to the fundamental property of quantum entanglement, which
cannot be explained by any classical theory. A consistent
discussion of the thermodynamics of systems that are monitored by
quantum information processing devices should therefore be based on an
appropriate definition of the algorithmic complexity of quantum states. 

In this paper, we propose a definition for the algorithmic complexity
of a quantum 
state that depends on a given precision parameter, and
give upper bounds on complexity for various examples of states. We
also establish a connection 
between the entanglement of a quantum state (in terms of its Schmidt
measure \cite{hans}) and its algorithmic complexity. Earlier proposals
for the algorithmic complexity exist, which are based either on the
reproducibility of a quantum state via Turing machines
\cite{vitanyi,vandam} or on universal probability \cite{gacs}. Our
proposal is 
based on the idea that each quantum 
state is ultimately identified with an experimental preparation
process \cite{bohr}. The definition of the algorithmic complexity of
a quantum state is thus naturally reduced to the 
description complexity of its (abstract) preparation process. A
similar approach has been introduced in \cite{yamakami} in the
analysis of relations between computation complexity and
entanglement. 

To motivate our definition of the algorithmic complexity of a quantum
state, we consider the following scenario. Alice has created a
certain quantum state in her laboratory and wants to describe this
state to Bob, who is supposed to reproduce it in his laboratory. How 
difficult is it 
for Alice to describe to Bob the state of her system? In order to
answer this question we may distinguish the two situations
in which Alice and Bob communicate via a classical or a quantum
channel. In the latter case,
Alice may send to Bob the quantum state altogether (or in
some Schumacher compressed form) \cite{note1intro}. If the
communication  is classical, this is not possible. On the other hand,
as a quantum state can be regarded as the result of 
some experimental preparation procedure, Alice may choose to send to
Bob the latter. In this case, the complexity of a quantum state
is identified, in a very natural way, with the description complexity
of an experimental preparation procedure \cite{note2}. 
 
Even though the first approach may seem ``more quantum'', it lacks of an
important feature that we usually associate to the \emph{description}
of an object.  Even if Bob has the state sent by Alice,
he might not know what state he has received. In this paper, thus, we
shall follow the second approach and identify the algorithmic
complexity of a quantum state with its preparation 
complexity i.e. the  classical description complexity of the
preparation procedure.

To be able to communicate, Alice and Bob must first have agreed on a 
common language which they
are using to describe their preparation procedure. Ideally, they will
also use the same ``toolbox'' to compose their experiments and the
same words when referring to elements of this toolbox. The toolbox is
in general an abstraction from any real experimental scenario: in
quantum information theory such an
abstraction leads to defining the toolbox by a set of elementary
operations on a Hilbert space with a given tensor-product structure
and dimension. In particular such toolbox will include
the ability to prepare some standard reference state, and a finite set
of elementary 
operations. A complete preparation procedure is then
described as a sequence of unitary transformations and possibly
measurements applied to the reference state.

In classical information theory, the algorithmic complexity of an
objects measures the amount of information necessary
to reproduce it. The Kolmogorov complexity $K_\cl$ of a binary string
$\bomega$ is defined as \emph{the length of the shortest program that,
  running on a universal Turing machine, gives $\bomega$ as output.}
We stress that algorithmic complexity takes in account only the
length of the program (and thus the length of 
the description of the object) and not the time needed by the computer
to actually run the program. This last quantity is instead studied by
the computational complexity and is related to a different property of 
objects, that is their logical depth \cite{bennettlogical}.

It is always possible to reproduce a
string  $\bomega=\omega_{i_1}\omega_{i_2}\cdots$ by means of a program
of the form: ``\emph{write $\omega_{i_1}$
  $\omega_{i_2}\cdots$}''. This implies that the length of a 
string constitutes 
(up to a constant) an upper bound for the complexity of the
string itself: $K_\cl(\bomega_n)\lesssim
l(\bomega_n) = n$ \cite{notelesssim}. Naturally there are
sequences for which this 
upper bound is far too large: the complexity of a periodic string,
for example, grows only logarithmically 
with the length of the sequence. A string is said to be \emph{complex}
(or structureless, or random) if its algorithmic complexity grows
linearily with its 
length: these are the strings typically generated by random sources
(e.g. a coin toss).

Considering that  a
quantum state can be characterized by a sequence of elementary
operations (represented e.g. as a ``circuit''), we
define the complexity of a state referring 
to that of the circuit itself. A finite set of
gates (constituting a \emph{complete gate basis}) is suitable to
prepare any state up to an arbitrary precision. Through an
adequate \emph{coding}, the 
circuit is reduced to a (classical) string whose Kolmogorov
complexity is well defined and which can be associated to the original
state. In this way the algorithmic complexity of a state satisfies the
intuitive idea of complexity as a measure of ``how difficult'' it is
to prepare a state. Since with a finite number of gates
only a countable set of states can be prepared exactly,  it is
necessary to introduce a \emph{precision parameter} in such
definition. 

From now on we represent with $\mcq_N$ the space generated by $N$
qubits and with $|0\rangle_N \in \mcq_N$ the vector $|0\rangle_N =
|0\rangle^{\otimes N}$, where each $\snul$ is an element of the
computational basis $\{|0\rangle, |1\rangle\}$ of a single-qubit
Hilbert space.
We represent with $\mcc|0\rangle_N$ the result of the application
of a circuit $\mcc$ on the state $\snul_N$; if $\vert \langle \phi | \mcc
\snul_N\vert^2 \geq 1 - \epsilon$ we will say that $\mcc$ prepares
$|\phi\rangle$ with precision $\epsilon$ (where $0\leq\epsilon \leq
1$).  We say that two states $\sphi$ and $\spsi$ are
\emph{$\epsilon$-distinguishable} if $|\langle\psi\sphi|^2\leq
1-\epsilon$.

Once we fix a
complete gate basis $B$ and a code $\Omega$, the procedure to
compute the algorithmic complexity of state $\sphi$ is the following.\\
1. With the gates contained in the basis $B$, build a circuit
  $\mcc^{B,\epsilon}$ that prepares $\sphi$ with precision
  $\epsilon$.\\
2. Code the circuit, obtaining a classical sequence 
  $\bomega^\Omega(\mcc^{B,\epsilon}) = \omega^\Omega_{i_1}
  \cdots\omega^\Omega_{i_m}$ of symbols
  $\omega^\Omega_{i_k}=\omega^\Omega_{i_k}(\mcc^{B,\epsilon})\in\Omega$~. \\ 
The algorithmic complexity of a state, relative to the  
  basis $B$, the code $\Omega$ and the circuit $\mcc^B$, with
  precision parameter $\epsilon$ is:
  $K_\net^{\Omega,B,\mcc^{B,\epsilon}}(\sphi) =
  K_{\textrm{Cl}}(\bomega^\Omega(\mcc^{B,\epsilon}))~.$ \\
3. In general there are more circuits that prepare the same state
  $\sphi$, and in principle the corresponding complexities can
  be different. In order to define a property of the
  state itself (and not related  to the circuit used to reproduce it)
  we minimize over all of them.\\
{\bf{Definition.}}  The algorithmic complexity of the state $\sphi$,
  relative to the code 
    $\Omega$ and the gate basis $B$, with precision parameter
    $\epsilon$ is: 
  \beq
       K_{\net}^{\Omega,B,\epsilon}(\sphi) =
    \min_{\mcc^{B,\epsilon}\in\tilde\mcc^{B,\epsilon}} 
    K_{\net}^{\Omega,B,\mcc^{B,\epsilon}}(\sphi)
  \eeq
  where $\tilde\mcc^{B,\epsilon}$ is the set of all the circuits built
  with gates 
  from $B$ that prepare $\sphi$ with precision $\epsilon$.

Considering that the \emph{choice of code} and
\emph{basis} is arbitrary, it is necessary to study how it
influences the complexity of the state. It is also relevant to consider
how the complexity of a quantum state depends on the \emph{precision}
with which we are required to reproduce it.

{\bf{Dependence on the code.}}
If $\Omega$ and $\underline{\Omega}$ are two different codes, then, for
any state $\sphi$ and any precision parameter $\epsilon$:
$K_{\net}^{\Omega,B,\epsilon}(\sphi) =
K_{\net}^{\underline{\Omega},B,\epsilon}(\sphi) +
k_{\Omega,\underline{\Omega}}~,$ 
where $k_{\Omega,\underline{\Omega}}$ is a constant that depends only
on $\Omega$ and $\underline{\Omega}$. 
$k_{\Omega,\underline{\Omega}}$ is the length of a ``dictionary''
with which it is 
possible to translate the description made using one code to that made
using the other. Since both codes are finite, such dictionary is
finite too and, in the limit of big values of
$K_{\net}^{\Omega,B,\epsilon}(\sphi)$, its contribute is negligible. 
Considering this code-invariance property we can omit
explicitating the dependence on the code (we can
imagine fixing it once and for all) and write simply:
$\displaystyle{K^{B,\epsilon}_\net(\sphi)}$~. 

{\bf{Complexity and precision.}} Let us consider now how the
complexity of a state depends on the precision parameter.
It has been shown \cite{prl} that using only the 
set of all 1-qubit gates, plus the controlled not, it is
possible to reproduce any unitary operation $U$ over $\mcq_N$ using
$\mco\left(4^N\right)$ gates. However, if one is interested in
reproducing the action of a unitary operation on one particular (given)
state, only $\mco\left(2^N\right)$ such gates are sufficient. Thus, to
prepare any state $\sphi$ from the given initial state $\snul_N$ we
need at most  $\mco\left(2^N\right)$ gates. We consider now the
Solvay-Kitaev theorem \cite{kitaev} which implies 
that any circuit acting on $\mcq_N$ built with $m$ 1-qubit and $\cnot$
gates and can be reproduced up to precision $\epsilon$ using 
$\mco\left(m \log^c\left(\frac{m}{\epsilon}\right)\right)$
gates from a finite gate basis ($c \in [1,2]$ is a constant whose
exact value is not yet known). It follows that the action
of any unitary transformation on 
$\snul_N$ can be implemented (and thus any $\sphi\in\mcq_N$ can be
prepared) up to precision $\epsilon$ via a circuit built with
gates from any finite and complete basis; futhermore, the number of
gates in such a circuit is  $M\sim 2^N\log\frac{1}{\epsilon}$.

The length of the string that codes the circuit grows
linearly with the number of gates of the circuit itself: to
code a circuit that prepares a general state $\sphi\in\mcq_N$ we need
thus a sequence whose length is (proportional to) $M$. From what seen
above we then have:
\beq
K_\net^{B,\epsilon}(\sphi\in\mcq_N)\lesssim 2^N\log\frac{1}{\epsilon}~.
\label{UpperBound}
\eeq

We notice that this bound is much higher than the classical one, where
the complexity grows at most linearly 
with the number of bits of the string. Intuitively this reflects the
fact that the space of quantum objects is much richer 
than that of classical ones. In the last part of this work we will
also see that such difference can be explained by
quantum entanglement.

{\bf{Dependence on the basis.}} The definition of the algorithmic
complexity of a state has a dependence on the choice of the
basis. Given any state $\sphi$ there always exists a particular
basis using which the preparation of $\sphi$ is trivial. If Alice wants to
describe to Bob a state and  they have previously agreed on using a
certain gate basis, then Alice has only to describe the circuit
(passing the sequence $\bomega_\mcc$). If they have not
agreed on a particular gate basis, then Alice could indeed build a
circuit using the ``best'' basis, but in this case she would have to
describe the basis itself to Bob, and this would be a similarly
difficult task \cite{note4}.  

One might nevertheless wonder whether there is some particular basis 
that can describe all (or almost all) states with simple
circuits. If such a basis existed it would obviously be convenient for
Alice and Bob to agree on using that and (almost) all states would be
non-complex. In the following we
show that such a basis cannot exist as, once \emph{any} gate
basis is fixed, the number of non-complex states is small in
relation to the total number of states. 

In classical information theory it is known that the number of
compressible (bit) strings is ``small''; more precisely the ratio
between the strings with complexity smaller than a constant $k$
and the total number of $n$-bit strings is bound by $(2^k-1)2^{-n}$~. 

As we have seen in the previous
paragraphs, once we fix a basis $B$ and a
precision parameter $\epsilon$, we can associate to every quantum
state $\sphi$ a $(2^N\log\frac{1}{\epsilon})$-bit string
$\hat\bomega_\epsilon^B(\sphi)$ whose classical algorithmic complexity
coincides with the complexity  of $\sphi$.
The result seen above for classical bit strings can thus be
generalized to quantum states and one finds that the ratio between
compressible quantum states (such that $ K_\epsilon^B(\sphi)<k$) and
the total number of $\epsilon$-distinguishable normalized states is
bound by $2^{2^N\log\epsilon}(2^k-1)\simeq2^{2^N\log\epsilon+k}$.
Such a relation holds true also in the case when $k=k(N,\epsilon)$ is
a function. 
State $\sphi$ will be non-complex only if its complexity is
$o(2^N\log\frac{1}{\epsilon})$: this means that the right member of
the inequality
becomes:$2^{2^N\log\epsilon+o(2^N\log\frac{1}{\epsilon})}\sim
2^{2^N\log\epsilon}\ll  1$. Thus, one obtains that for any fixed
basis $B$, the number of non-complex states is exponentially small \cite{note5}.

Note that there are cases in which the complexity is invariant for
basis choice. This happens, for example, when we consider 
a \emph{coarsening} of the gate basis, that is if we consider two gate
bases $B$ and $\underline{B}$, one of which
($\underline{B}$) constituted of gates  that can be
built with gates from $B$ (e.g. $\underline{B}$ containes a Toffoli
gate, while $B$ contains Hadamard and C-not). In this case, 
any circuit made by gates from $\underline{B}$ can be reproduced by
one made by gates from $B$. The string that codes this circuit will in 
general be longer than that of the original circuit, but their
complexities will change only for a (small) constant
$k_{B,\underline{B}}$ (that represents the length of a ``dictionary''
between the two gate bases). \\
From a less abstract point of view, this property translates into a
form of invariance with respect of the choice of the experimental
apparatus. It is not required that the circuit be actually
built only with the elementary gates: the use of more complex
components does not modify the complexity of the description as long
as they are themselves composed of elementary parts.

{\bf{Entanglement and complexity.}} The nature of quantum
correlations has been a central issue of long-lasting debates on the
interpretation of quantum mechanics.In the last years, the notion of
entanglement has been recognized as central to quantum information
processings \cite{nielsen}.  As a result, the task of characterizing quantum entanglement, and the properties related to it, has emerged as
one of the prominent themes of quantum information theory. 
In the last part of this letter we investigate the relations 
between the algorithmic complexity of a state and its
entanglement properties.

Let us begin by considering the case of a $J$-separable state $\sphi\in\mcq_N$ of
the type: $\sphi =\bigotimes_{j=1}^J|\phi_j\rangle$, 
with $|\phi_j\rangle\in\mcq_{N_j}$, dim$(\mcq_{N_j})=2^{N_j}$, and
$\sum_{j=1}^J N_j=N$ .
State $\sphi$ is not totally entangled, as it can
be written as the tensor product of other (possibly entangled)
states $|\phi_j\rangle$ (with
$K_\net^{B,\epsilon}(|\phi_j\rangle)\leq 2^{N_j}\log\frac{1}{\epsilon}$). As
a consequence of this fact we have: \mbox{ 
$K_\net^{B, \epsilon}(\sphi) \leq
\sum_{j=1}^JK_\net^{B,\epsilon/J}(|\phi_j\rangle)
<2^N\log\frac{1}{\epsilon}$},
where we have used the fact that $\sphi$ is prepared with precision
$\epsilon$ if all states $|\phi_j\rangle$ are
prepared with precision $\epsilon/J$ \cite{noi}. Given any $J\geq 1$, thus, there exist $(J-1)$-separable states whose complexity is larger than that of \emph{any} $J$-separable state.

Thus, the {\emph{maximal complexity}} can be obtained {\emph{only}} by a
truly $N$-party entangled state (in the sense that it cannot be written as
tensor product of states contained in subspaces of $\mcq_N$). We
stress that this consideration does not imply that all totally
entangled states have maximal complexity. Counterexamples are given by W, GHZ, and $N$-qubit graph states \cite{WGHZ}. The latter 
are highly entangled, but nevertheless their complexity is bound by
$N^2\log\frac{1}{\epsilon}$ \cite{noi}. As another example let us
consider a state $\sphi\in \mcq_N$ of the form $\sphi =
\bigotimes_{j=1}^N  |\phi_j\rangle,~~|\phi_j\rangle\in\mcq_1$. In this
case we have: $K_\net^{B,\epsilon}(\sphi)\leq\sum_{j=1}^N
K_\net^{B,\epsilon/N}(|\phi_j\rangle) \leq 2 N
\log\frac{N}{\epsilon}$. Thus the complexity of a separable state can 
grow only at most linearly with the number of qubits. 

The relation, illustrated  by the  examples above, between
complexity and entanglement can be formalized if we
choose the \emph{Schmidt measure} \cite{hans} as a measure of
entanglement.  Given a quantum state $\sPhi\in\mcq_N$, its Schmidt
measure $E_S(\sphi)$ is defined as the logarithm of the minimum number 
$r$ of separable states $|\phi_i\rangle\in\mcq_N$ such that $\sPhi =
\sum_{i=1}^r \alpha_i|\phi_i\rangle\in\mcq_N$. In the following, we
show how the knowledge of the Schmidt 
measure of a quantum state $\sPhi$ allows us to give an upper bound on 
its complexity. 

In order to do this we specify a circuit that
prepares $\sPhi$. The general idea is to compose such a circuit
``using'' the circuits that prepare the states $|\phi_i\rangle$ that
appear in the decomposition of $\sPhi$. Such a circuit is built with
the aid of $\log r$ ``ancilla'' qubits, initially prepared in the
superposition state $|a\rangle=\sum_{i=0}^r \alpha_i|i\rangle$
(where $|i\rangle$ is the state whose binary representation gives the
state of the $\log r$ qubits). We then apply to the initial state
$\snul$ a circuit consisting of 
controlled-$\mcc^B_{\epsilon}(|\phi_i\rangle)$ (in series), the
application of each conditional on the 
ancilla being in state $|i\rangle$. The last step is to project the
ancilla on state $\sum_{i=1}^r
|i\rangle/\sqrt{r}=\bigotimes_{j=1}^{\log r}(\snul_j +
|1\rangle_j)/\sqrt{2}$: state $\sPhi$ is prepared 
if the result is non-zero. If this is not the case, it is sufficient
to repeat the procedure from the beginning \cite{NoteSchmidtCircuit}.
The complexity of the state $\sPhi$ can now be expressed in terms of 
the complexity of the ancilla state $|a\rangle$ and of that of the
remaining circuit.

The ancilla can in principle be any state of $\log r$ qubits; from
what was seen in the above sections we have thus:
$K_\net^{B,\epsilon}(|a\rangle) \leq 2^{\log r}\log\frac{1}{\epsilon} =
r\log\frac{1}{\epsilon}$. Once the ancilla is prepared in the required
state, it is necessary to describe the remaining circuit. First of all
we must describe the individual circuits
$\mcc^B_{\epsilon}(|\phi_i\rangle)$ (and this requires a number of
bits bounded by the sum of the complexities 
of the single states $|\phi_i\rangle$). Performing each of these
circuits as a controlled operation requires additional $\mco(\log
r \log\frac{1}{\epsilon})$ gates from $B$. We thus have:
\beq
K_\net^{B,\epsilon}(\sPhi)\lesssim 3N
2^{E_S(\sPhi)}\log\frac{1}{\epsilon}~, 
\label{entbound}
\eeq
where we have used the fact that, as the states $|\phi_i\rangle$ are 
separable, their complexity is bound by $2N\log\frac{1}{\epsilon}$.\\
Note that this result includes the two particular cases
treated previously. If $\sPhi$ is separable, in fact, we have
$E_S(\sPhi)=0$, and the equation above takes the form:
$K_\net^{B,\epsilon}(\sPhi) \leq
3N\log\frac{1}{\epsilon}$. Alternatively, when the 
state $\sPhi$ is truly entangled, with Schmidt measure 
$\log r=N$, we have $K_\net^{B,\epsilon}(\sPhi) \leq
2N 2^N\log\frac{1}{\epsilon}\sim 2^N\log\frac{1}{\epsilon}$ (up to a
polynomial term).  

The notion of complexity descibed in this letter can be
extended to mixed quantum states. Given a state $\rho$
its algorithmic 
complexity is defined as $K_\net^{B,\epsilon}(\rho) =
\min \prod_i [K_\net^{B,\epsilon}(|\phi\rangle)]^{\lambda_i}$, where the
minimum is taken over all pure state mixtures such that
$\rho=\sum_i\lambda_i|\phi_i\rangle\langle\phi_i|$. Thus the
complexity of a mixed state is the average complexity of the states
that appear in its (optimal) mixture. The choice of the geometric
average is dictated by the necessity of giving an adequate weight to
the probability distribution. Such a definition coincides with the one
we have introduced in this paper in the pure state case. Furthermore one
finds that the relationship between entanglement and complexity also
holds. In particular we have: $K_\net^{B, \epsilon}\lesssim 3N
2^{E_S(\rho)}\log\frac{1}{\epsilon}$, where $E_S(\rho)$ is the Schmidt
measure of the mixed state $\rho$ as defined in \cite{hans}.

In this paper we have introduced a new notion of algorithmic
complexity of a quantum state, based on the classical description of
its preparation procedure. We have seen how the complexity of a
quantum state grows, in general, exponentially with the number of
qubits and this is significantly different from the algorithmic
complexity of classical objects (where the upper bound is linear with
the number of bits). In the last part we have shown how this
difference can be interpreted as a consequence of the presence of
quantum correlations, by giving a bound on the complexity of a state
in terms of its entanglement. A consequence of this is that the absence
of entanglement (i.e. in completely separable states) re-establishes
the classical limit for the complexity bound. Entanglement thus proves
again being a fundamental feature that distinguishes quantum objects
from classical ones. 

We wish to thank M. Bremner and M. Piani for their help in editing the
manuscript. This work was
supported in part by the FWF, the DFG, and the EU (IST-2001-38877,
-39227, OLAQUI, SCALA).


\begin{thebibliography}{99}


\bibitem{solomonoff} R. J. Solomonoff, Information and Control
  {\bf{7}}, 1 (1964).

\bibitem{kolmogorov} A. N. Kolmogorov, Probl. Inf. Transmission
  {\bf{1}}, 1 (1965).

\bibitem{chaitin} G. J. Chaitin, \emph{Information, Randomness \&
  Incompleteness} (World Scientific, 1987).

\bibitem{bennett} C. H. Bennett, Internal J. Theoret. Phys. {\bf{21}}, 905
    (1982).

\bibitem{zurek} W. H. Zurek, Nature {\bf{341}}, 119 (1989).

\bibitem{schack} R. Schack and C. M. Caves, Phys. Rev. Lett. {\bf{69}}
  (1992).

\bibitem{caves} C. M. Caves, Phys. Rev. E {\bf{47}} (1993).

\bibitem{hans} J. Eisert and H. J. Briegel, Phys. Rev. A
    {\bf{64}} (2001).

\bibitem{vandam} A. Berthiaume et al.,
  J. Compute. System Sci. {\bf{63}} (2001).

\bibitem{vitanyi} P. Vitanyi, IEEE Trans. Inf. Theory {\bf{47}}, 2464
  (2001).

\bibitem{gacs} P. G\'acs, J. Phys. A {\bf{34}}, 6859 (2001).

\bibitem{bohr} N. Bohr, \emph{Atomic Physics and Human Knowledge}
  (Wiley, New York, 1958).

\bibitem{yamakami} T. Yamakami,  arXiv: quant-ph/0412172 (2003).

\bibitem{note1intro} Adopting this scenario we arrive to a notion of
  complexity similar to that introduced in \cite{vandam}.

\bibitem{note2} This
 approach is close to Bohr's viewpoint that the quantum state is
 essentially  an expression of an experimental scenario \cite{bohr}.  

\bibitem{bennettlogical} C. H. Bennett, \emph{How to Define Complexity
  in Physics and Why} (Addison-Wesley, Redwood City California, 1990),
  pp. 137-148.

\bibitem{notelesssim} Here and in the following we use the
   signs ($\simeq,~\lesssim$) when the relationships
  are expressed up to the leading order. 

\bibitem{prl} J. J. Vartiainen et al.,
  Phys. Rev. Lett {\bf{92}}, 177902 (2004).

\bibitem{kitaev} A. Y. Kitaev, Russ. Math. Surv. {\bf{52}}, 1191 (1997).

\bibitem{note4} This, in fact, would again require
  to describe arbitrary unitary transformations.

\bibitem{note5} Even though almost all states are complex, it is impossible to prove that any given state is. This derives from the fact that the statement:``The complexity of the classical string $\bomega$ is greater than $N$'' is unprovable \cite{chaitin_scient}.

\bibitem{chaitin_scient} G. J. Chaitin, \emph{Randomness and Mathematical Proof}, Scientific American {\bf{232}} (5) (1975). 

\bibitem{nielsen} M. A. Nielsen and I. L. Chuang, \emph{Quantum
    Computation and Quantum Information} (Cambridge University Press,
    2000). 

\bibitem{noi} C. E. Mora and H. J. Briegel, arXiv: quant-ph/0412172.

\bibitem{WGHZ} By W and GHZ states we intend $N$/qubit entangled states of the form, respectively, $(|0...01\rangle + |0...10\rangle + ... + |1...00\rangle)/\sqrt{N}$ and $(|00...0\rangle + |11...1\rangle)/\sqrt{2}$.

\bibitem{NoteSchmidtCircuit} This procedure succeeds with
  exponentially low probability, 
but this is not relevant from our point of view: we are in fact
interested in the description of the procedure, not in the time or
resources needed to apply it.

\end{thebibliography}

\end{document}